\documentstyle[prd,aps,a4]{revtex}
\input epsf

\newcommand\lsim{\mathrel{\rlap{\lower4pt\hbox{\hskip1pt$\sim$}}
    \raise1pt\hbox{$<$}}}
\newcommand\gsim{\mathrel{\rlap{\lower4pt\hbox{\hskip1pt$\sim$}}
    \raise1pt\hbox{$>$}}}
\newcommand\esim{\mathrel{\rlap{\raise2pt\hbox{\hskip0pt$\sim$}}
    \lower1pt\hbox{$-$}}}

\begin{document}
\title{Can we predict the fate of the Universe?}

\author{P. P. Avelino${}^{1,2}$\thanks{
Electronic address: pedro\,@\,astro.up.pt},
J.\ P.\  M.\ de Carvalho${}^{1,3}$\thanks{
Electronic address: mauricio\,@\,astro.up.pt}
and C. J. A. P. Martins${}^{4}$\thanks{Also at C.A.U.P.,
Rua das Estrelas s/n, 4150 Porto, Portugal.
Electronic address: C.J.A.P.Martins\,@\,damtp.cam.ac.uk}}

\address{${}^1$ Centro de Astrof\'{\i}sica, Universidade do Porto\\
Rua das Estrelas s/n, 4150 Porto, Portugal}

\address{${}^2$ Dep. de F\'{\i}sica da Faculdade de Ci\^encias da Univ. do Porto,\\
Rua do Campo Alegre 687, 4169-007 Porto, Portugal}

\address{${}^3$Dep.\ de Matem\'atica Aplicada da Faculdade de Ci\^encias da Univ.\ do Porto,\\
Rua das Taipas 135, 4050  Porto, Portugal}

\address{${}^4$ Department of Applied Mathematics and Theoretical Physics\\
Centre for Mathematical Sciences, University of Cambridge\\
Wilberforce Road, Cambridge CB3 0WA, U.K.}

\maketitle
\begin{abstract}
We re-analyze the question of the use of cosmological observations
to infer the present state and future
evolution of our patch of the universe. In particular,
we discuss under which conditions one might be able to infer that our
patch will enter an inflationary stage, as a {\em prima facie}
interpretation of the Type Ia supernovae and CMB data would suggest.
We then establish a `physical' criterion for the existence of inflation, to be
contrasted with the more `mathematical' one recently proposed by Starkman
{\em et al.} \cite{STV}.
\end{abstract}
\pacs{98.80.Es, 98.80.Cq, 98.62.Py\\
Keywords: Gravitation; Cosmology; Inflation; Observational Tests}

\section{\bf Introduction}
\label{intro}
The issue of the present state,
future dynamics and final fate of the Universe, or at least our patch
of it, has been
recently pushed to the front line of research in cosmology. This is mostly
due to observations of high redshift type Ia supernovae, performed by
two independent groups (the
``Supernova Cosmology Project'' and the ``High-Z Supernova Team''), which
allowed accurate measurements of the luminosity-redshift relation out to
redshifts up to about $z \sim 1$ \cite{Perlmu,Riess1,Garnavich}.
It should be kept in mind that these
measurements are done on the assumption that these
supernovae are standard candles, which is by no means demonstrated and
could conceivably be wrong.
There are concerns about the evolution of these objects
and the possible dimming caused by intergalactic dust \cite{Drell,Riess2}, but
we will ignore these for the purposes of this paper, and assume that the
quoted results are correct.

The supernovae data, when combined with the ever growing
set of CMBR anisotropy observations,
strongly suggest an accelerated expansion of the Universe
at the present epoch, with cosmological parameters
$\Omega_\Lambda \sim 0.7$ and $\Omega_{\rm m} \sim 0.3$. A further cause of
concern here is the model dependence of the CMBR
analysis, but we shall again accept the above results for the
purpose of this paper.

Taken at face value, these results would seem to show that the universe 
will necessarily enter an inflationary stage in the near future. 
However, as pointed out by Starkman, Trodden and
Vachaspati \cite{STV} this is not necessarily so. We could be living in a
small, sub-horizon bubble, for example. And even if we were indeed inflating,
it would not be trivial to demonstrate it. 
In the above work, these authors looked
at the crucial question of `How far out must we look to infer that
the patch of the
universe in  which we are living is inflating?' Their analysis is based on
previous work by Vachaspati and Trodden \cite{VT} which shows that the onset of
inflation can in some sense be identified with the comoving contraction of
our minimal anti-trapped surface (MAS)\footnote{The MAS of each comoving
observer is a sphere centered on him/her, on which the velocity of comoving
objects is $c$. For the particular case of an homogeneous universe, the MAS
has a physical radius $cH^{-1}$.}.
They then argue that if one can confirm cosmic acceleration
up to a redshift $z_{MAS}$ and detect the contraction of our MAS, then our
universe must be inflating. Unfortunately, even if we can do the former (for
$\Omega_\Lambda=0.8$ the required redshift is $z_{MAS}\sim1.8$), it turns
out that there is no way to presently
confirm the latter, because the accelerated
expansion hasn't been going for long enough for the MAS to contract. Only if
we had $\Omega_\Lambda\ge0.96$ would we be able to demonstrate inflation
today.

As in the proverbial mathematicians joke, the method outlined by Starkman
{\em et al.} provides an answer that is completely accurate but will take a
long time to find, and hence is of no immediate use to us. In this paper, 
however, we will explore a different possibility.
Our main aim is to provide what could be called a physicists
version of the ``mathematical'' question of Starkman {\em et al.} \cite{STV}.
In other words, we are asking, {\em `If we can't know for sure
the fate of the universe at
present, what is our best guess today?'}. As we will discuss, we can answer
this question, although it will involve making
some crucial additional assumptions.

In order to answer the above question we compare the particle and event
horizons. We show that for a flat universe with $\Omega_\Lambda \gsim 0.14$
the particle horizon is greater than the distance to the 
event horizon meaning that today
we may be able to observe a larger portion of the Universe than that which
will ever be able to influence us. We argue that if we find evidence for
a constant vacuum density up to a distance from us equal to the event
horizon then our Universe will necessarily enter an inflationary phase
in the not too distante future, {\em assuming} that the potential 
of the scalar field which
drives inflation is time-independent and that the content of the
observable universe will remain `frozen' in comoving coordinates.

Note that Starkman {\em et al.}
argue that inflation can only take place if the vacuum energy dominates
the energy density on a region with physical radius not smaller than
that of the MAS at that time. However, they did not assume that the 
content of the observable universe would remain frozen in comoving
coordinates and so they found that the larger is the contribution of a
cosmological constant for the total density of the Universe, the
larger is the redshift out to which one has to look in order to infer
that our portion of the universe is inflating.
This result seems paradoxical, until one realizes that the size of the MAS
at a given time {\em does not}, by itself, say anything about inflation.
The main reason why $z_{MAS}$ grows with $\Omega_\Lambda$ is
simply because the scale factor has grown more.

The plan of this paper is as follows. In the next section we introduce the
various lengthscales that are relevant to our discussion, and
provide a qualitative discussion of our
test for inflation. We also discuss the assumptions involved and compare our
`physical' test with the `mathematical' one recently proposed by
Starkman {\em et al.} \cite{STV}. In section \ref{biaf} we provide a more
quantitave analysis of our criterion. We also discuss in more detail our 
crucial assumption of an energy-momentum distribution which remains 
frozen in comoving coordinates. Finally, in section \ref{conc} we 
summarize our results and discuss some other outstanding issues.

\section{A `physical' test for inflation}
\label{horizon}
The dynamical equation which describes the evolution of the scale factor
$a$  in a Friedmann-Robertson-Walker (FRW) universe containing matter, 
radiation and a cosmological constant can be written as
\begin{equation}
 H^2=H_0^2(\Omega_{m0} a^{-3} + \Omega_{r0} a^{-4} +
\Omega_{\Lambda0} + \Omega_{k0} a^{-2}).
\label{one}
\end{equation}
where $H={\dot a / a}$ and the density parameters $\Omega_m$, $\Omega_r$ and
$\Omega_\Lambda$ express respectively the densities in matter, radiation and
cosmological constant as fractions of the critical density\footnote{A dot
represents a derivative with respect to the cosmic time $t$. The
subscript `$0$' means that the quantities are to be evaluated at
present epoch, and we have also taken $a_0=1$.}.
Naturally one has $\Omega_k = 1 - \Omega_m-\Omega_r-\Omega_\Lambda$.

The distance $d$, to a comoving observer at a redshift $z$ is
given by
\begin{equation}
 d(z)= c \int_{t(z)}^{t_0} {{d t'} \over a(t')} = c H_0^{-1}
\int_0^z {{dz'} \over { \left[ \Omega_{m0} (1+z')^3 + \Omega_{r0} (1+z')^4 +
\Omega_{\Lambda0} + \Omega_{k0} (1+z')^{2}
\right]^{1/2}}} \, ,
\label{two}
\end{equation}
and is related to the `radius' of the local universe which we can in
principle  observe today. The distance to the 
{\em event horizon} can be defined as
\begin{equation}
 d_e= c \int_{t_0}^\infty {{d t'} \over a(t')} = c H_0^{-1}
\int_1^\infty {{da} \over {  (\Omega_{m0} a + \Omega_{r0} + 
\Omega_{\Lambda0} a^4 + \Omega_{k0} a^{2})^{1/2}}}.
\label{three}
\end{equation}
and represents the portion of the Universe which will ever be able to
influence us\footnote{In writing the upper integration limit as infinity we are
of course assuming that the universe will keep expanding forever; an
analogous formal definition could be given for an universe ending in a
`big crunch'.}. On the other hand,
the {\em particle horizon}, $d_p$, is defined by (from eqn. (\ref{two}))
\begin{equation}
d_p\equiv \lim_{z \to \infty} d(z)\, ,
\label{five}
\end{equation}
and it represents the maximum distance which we can observe today.

If today the distance to the event horizon is smaller than 
the particle horizon ($d_e <
d_p$) this means that today we are able to observe a larger portion of
the Universe than that which will ever be able to influence us.
We can do this if we look at a redshift greater than $z_*$ defined by
(see also \cite{staro})
\begin{equation}
d(z_*)=d_e\, .
\label{four}
\end{equation}
In a flat universe solutions to this equation are only
possible for $z_* \ge 1$ and for $\Omega_{\Lambda0} \gsim 0.14$.
Hence, assuming that the energy-momentum distribution within the 
patch of the Universe which we are able to see remains unchanged 
in comoving coordinates, our Universe will
necessarily enter an inflationary phase in the future if there is a
uniform vacuum density permeating the Universe up to a redshift
$z_*$.

This assumption obviously requires
some further discussion. One can certainly think of a
universe made up of different `domains', each with its own values of
the matter and vacuum energy
density. Furthermore, by cleverly choosing the field dynamics, one can
always get patches with time-varying vacuum energy densities, or patches
where the vacuum energy density is non-zero for only short periods.
In all such cases, the domain walls separating these
patches can certainly have a very complicated dynamics, and in
particular it is always
possible that a domain wall will suddenly get inside our horizon sometime
between the epoch corresponding to our observations and the present day.
On the other hand, it should also be pointed out that a certain amount of
fine-tuning would be required to have a bubble coming inside our horizon 
right after we have last observed it.
In these circumstances, the best that can be done is to impose constraints
on the characteristics of any bubble wall that could plausibly have entered
the patch of the universe we are currently able to observe, given that we
have so far seen none. We shall analyse this point in a more quantitative
manner in the following section.

\begin{figure}
\vbox{\centerline{
\epsfxsize=0.8\hsize\epsfbox{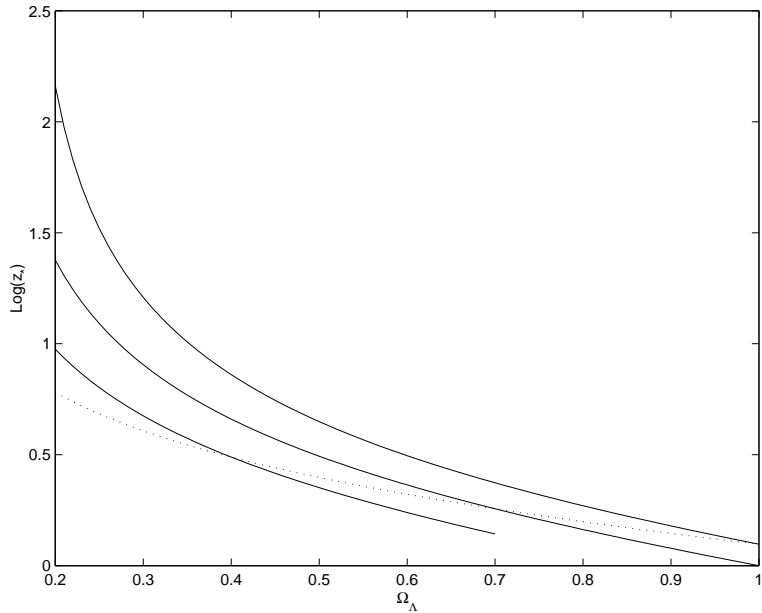}}
\vskip.2in}
\caption{The solution of eqn. (\ref{four}) for cosmologies with $\Omega_m+\Omega_\Lambda=0.7, 1.0, 1.3$ (solid curves, bottom to top)
and $\Omega_m=0.3$ for illustration (dotted curve). Observing a uniform
vacuum energy density up to a redshift $z_*$ will imply that our universe
will enter an inflationary phase in the future, subject to the conditions
specified in the text.}
\label{fig1}
\end{figure}

We think that the results obtained in this way, even if less robust from
a formal point of view, are intuitively more meaningful than
those obtained in \cite{STV} in the sense that, among other things,
in this case the minimum 
redshift $z_*$ out to which one must observe in order to be able to
predict an inflationary phase (subject to the conditions mentioned earlier)
decreases as $\Omega_\Lambda$ increases---see Fig. \ref{fig1}. In other
words, the larger the present value of the cosmological constant, the easier
it should be to notice it.

It is perhaps instructive to compare our test with that of \cite{STV}
in more detail. Starkman {\em et al.} require
the contraction of the MAS. Now, in order to see the MAS one
has to look at a redshift defined by:
\begin{equation}
a(z_{MAS})d(z_{MAS}) - c H^{-1}(z_{MAS})=0\, .
\label{six}
\end{equation}
This finds the redshift, $z_{MAS}$, for which the physical distance to a
comoving observer at that redshift, evaluated at
the corresponding time $t_{MAS}$, is equal to the Hubble radius at that time.
However, if the vacuum density already dominates
the dynamics of the Universe at the redshift $z_*$ then eqn.
(\ref{four}) reduces to:
\begin{equation}
d(z_*) - c H^{-1}(z_*)=0.
\label{seven}
\end{equation}
(recall that $a_0=1$)
because during inflation the physical size of the event horizon is 
simply equal\footnote{This is only exactly true when the vacuum energy
density is the only contributor to the energy density, in which case
exponential inflation occurs.}
to $cH^{-1}$  (this is ultimately the reason for the choice of 
criterium for inflation by Vachaspati and Trodden \cite{VT}). 
As has been discussed above, eqns. (\ref{six}) and (\ref{seven}) 
have totally different solutions ($z_*=1$ while $z_{MAS} \to \infty$).

To put it in another way,
the main difference between our approach and that of
ref.\cite{STV} lies on the fact that we assume that the energy-momentum
content of the observable Universe does not change
significatively in comoving coordinates. This allows us to use the 
equation of state of the local universe
observed for a redshift $z$ (looking back at a physical time $t(z)$) 
to infer the equation of state of the local universe at the present time.
In the following section, we shall discuss these points in somewhat
more detail.

We should also point out that if we were to relax the assumption of a
co-movingly frozen content of the observable universe, then the
equation---analogous to (\ref{four})---specifying the redshift out to which
one should look in order to be able to predict the future of the Universe
would be
\begin{equation}
d(z_+)=d_e(z_+)\, .
\label{fourtwo}
\end{equation}
This equation has no solution, so a stronger test of this kind is not feasible
in practice.

\section{\bf Discussion}
\label{biaf}

Here we go through some specific aspects of our test in more
quantitative detail. To begin with, we
have solved numerically eqn. (\ref{four});
the numerical results were obtained for choices of cosmological
parameters such that $\Omega_m+\Omega_\Lambda=0.7, 1.0, 1.3$,
with an additional $\Omega_m=0.3$ for illustration.
We are interested only in a matter--dominated or $\Lambda-$dominated
epoch of the evolution of the universe, and therefore we have dropped the
radiation density parameter $\Omega^{\rm r}_0$ of eqn. (\ref{one}), in the
calculations.

These results are displayed in Fig. \ref{fig1} as a function of
$\Omega_{\Lambda0}$. The cases with constant total density
are shown in solid curves (with the top curve corresponding to the higher
value of the density), while the case of a fixed  $\Omega_{\rm m}$,
is shown, for comparison purposes, by a dotted curve.

As expected, as the universe becomes more $\Lambda-$dominated and/or
less matter--dominated, the comoving distance to the event
horizon decreases, which is reflected in the decrease of the redshift
$z_*$ of a comoving source located at that distance\footnote{Note that
pushing $\Omega_{\Lambda0}$ down to zero, the value of $z_*$ tends
to infinity, since in such universes an event horizon does not exist.}.
For the observationally preferred values of $\Omega_m=0.3$
and $\Omega_\Lambda=0.7$, the required redshift will be $z_*\sim 1.8$.

\begin{figure}
\vbox{\centerline{
\epsfxsize=0.8\hsize\epsfbox{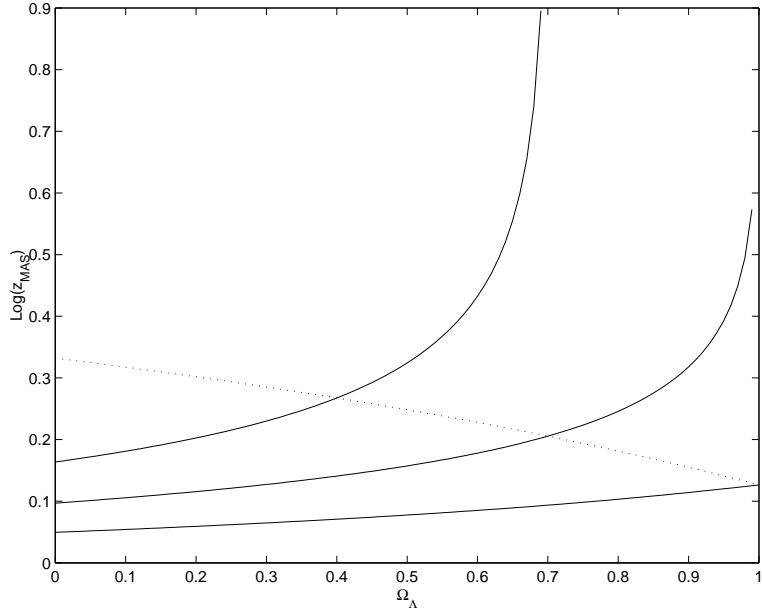}}
\vskip.2in}
\caption{The redshift of the MAS, relevant for the inflationary criterion
of Starkman {\em et al.}, for the same cosmological models as
in Fig. \ref{fig1}. Note that the solid curves
for $\Omega_m+\Omega_\Lambda=0.7, 1.0, 1.3$ now appear in the
graph from top to bottom.}
\label{fig2}
\end{figure}

\begin{figure}
\vbox{\centerline{
\epsfxsize=0.8\hsize\epsfbox{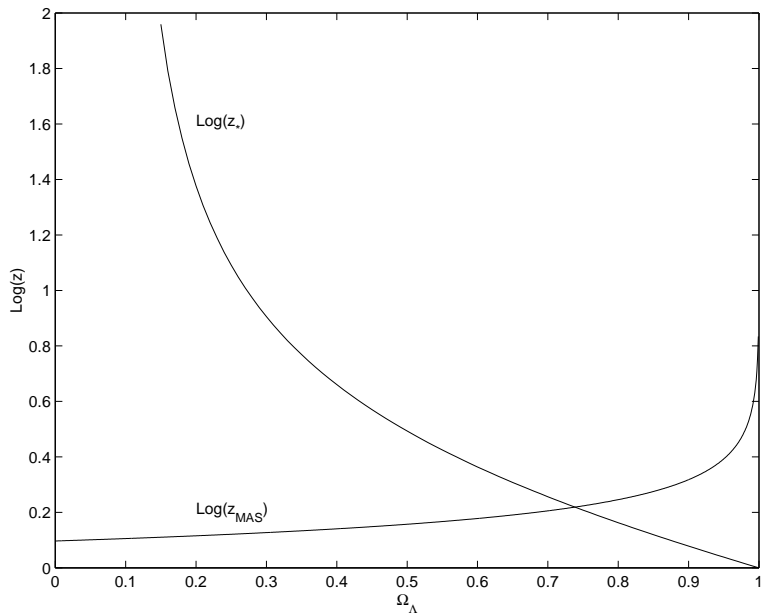}}
\vskip.2in}
\caption{Comparing the values of the critical redshifts $z_*$ and $z_{MAS}$,
as a function of the vacuum energy density,
for the spatially flat models ($\Omega_m+\Omega_\Lambda=1.0$).}
\label{FNEW}
\end{figure}

For comparison, the redshift, $z_{MAS}$, defined by eqn. (\ref{six}), 
which is the analogous relevant quantity
for the criterion of Starkman {\em et al.} is shown, for the same choices
of cosmological parameters, in Fig. \ref{fig2}. Note that in this case,
as the universe becomes more $\Lambda-$dominated and/or
less matter--dominated, the redshift of the MAS will increase.
As we already pointed out our test will not be applicable for very low
values of the vacuum energy density, and for intermediate values,
it requires a higher redshift than $z_{MAS}$.
However, for high values of the cosmological constant and/or low
matter contents, the fact that the universe will be expanding much faster
makes the redshift of the MAS increase significantly, and even become
larger than $z_*$ for some combinations of cosmological parameters.
For the same observationally preferred values of $\Omega_m$
and $\Omega_\Lambda$ quoted above,
the required redshift will be $z_{MAS}\sim 1.6$. A comparison of the
values of $z_*$ and $z_{MAS}$ for the spatially flat model is shown in
Fig. \ref{FNEW}.
 
We now return to our assumption about the 
energy-momentum content of the universe, considering the possibility that
different regions of space may have 
different values for the vacuum energy density, which are
separated by domain walls. This means that we are assuming the
existence of a scalar field, say $\phi$, which within each region
sits in one of a number of possible minima of a 
time-independent potential. It is obvious that if the potential 
depends on time or if the scalar field did not have time to roll to
the minimum of the potential, then it is not possible to predict the fate of
the universe without knowing more about the particle physics model which
determines its dynamics. For simplicity, we shall assume
that we live in a spherical domain with constant vacuum energy density 
(effectively a cosmological constant) that is
surrounded by a much larger region in 
which the the vacuum density has a different value---for the present 
purposes we will assume it to be zero. Note that this is the case where
the dynamics of the wall will be faster (more on this below). 

Is it possible that a region 
with a radius $d(z)$, say centred on a nearby observer,
can be inside a given domain at the conformal time $\eta(z)$,
but outside that domain at the present time, $\eta_0$? This
problem can provide some measure of how good the assumption of 
a frozen energy-momentum distribution in comoving coordinates is. 
In other words, is it likely that a domain 
wall may have entered this region at a redshift smaller than $z$?
In order to provide a more quantitative answer to this question,
we have performed numerical simulations of domain wall evolution using
the PRS \cite{PRS} algorithm, in which the thickness of the domain walls
remains fixed in comoving coordinates for numerical convenience.
See also \cite{AM} for a description of the simulations.

We assume that the domain wall has spherical symmetry, thereby
reducing a three-dimensional problem to a one-dimensional 
one. We perform simulations of this wall in a flat universe on a
one-dimensional $8192$ grid. The comoving grid spacing is $\Delta x=c \eta_i$
where $\eta_i=1$ is the conformal time at the beginning of the simulation. 
The initial comoving radius of the spherical domain was chosen to
be $R=2048 \Delta x$, and the comoving thickness of 
the domain wall was set to be $10 \Delta x$. In these simulations 
we neglect the gravitational effect induced by the different domains and
domain walls on 
the dynamics of the universe, and we also do not consider the possibility of
an open universe. We must emphasise, however, that both these effects would
slow down the defects, thereby helping to justify our assumption of a constant
equation of state in comoving coordinates even more.

We have obtained the following 
fit for the radius of the domain wall as a function of the conformal 
time $\eta$
\begin{equation}
R(\eta)=R_\infty\left(1-\left({{c \eta} \over {\alpha R_\infty}}\right)^n\right)^{2/n}\, ,
\label{fit1}
\end{equation}
where
\begin{equation}
\alpha=2.5\, , \qquad  n=2.1\,
\label{powers}
\end{equation}
and $R_\infty$ is the initial comoving radius of the domain wall 
(with $R_\infty \gg c \eta_i$). 
This fit is accurate to better than $5 \%$, except for the final stages of collapse.
In a flat universe with no cosmological constant
the comoving distance to a comoving object at a redshift $z$ is given by
\begin{equation}
d(z)=c \eta_0\left(1-1/\sqrt{1+z}\right)\, ,
\label{reldz}
\end{equation}
whereas the radius of the spherical domain wall can be written as
a function of the redshift $z$, given its initial radius $R_\infty$,
as follows
\begin{equation}
R(z)=R_\infty\left(1-\left({{c \eta_0} \over {\alpha R_\infty \sqrt{1+z} }}\right)^n\right)^{2/n}\, .
\label{reldz2}
\end{equation}

\begin{figure}
\vbox{\centerline{
\epsfxsize=0.8\hsize\epsfbox{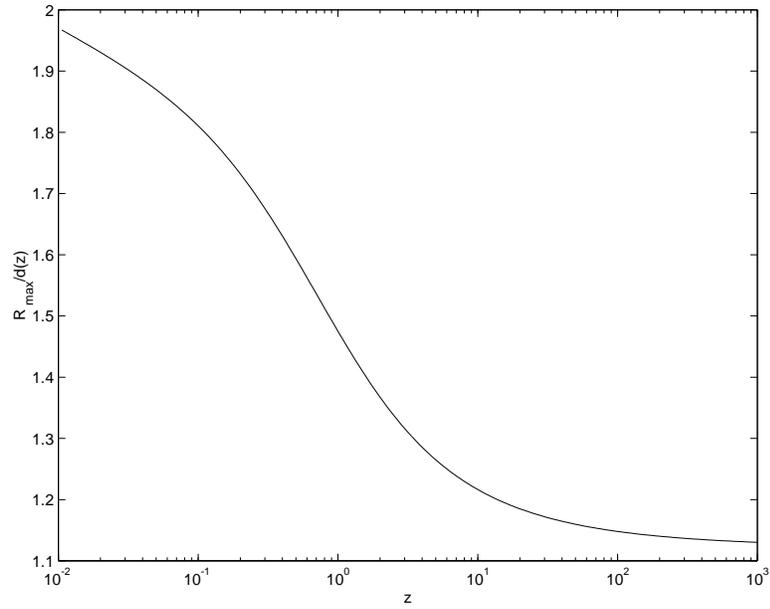}}
\vskip.2in}
\caption{The comoving radius of a domain wall at redshift $z$ whose present
comoving size equals the comoving distance to an object at
redshift $z$---denoted $d(z)$, see (\ref{reldz})---in units of $d(z)$,
as a function of redshift.}
\label{fig3}
\end{figure}

Now, by solving the equation
\begin{equation}
R(z=0)=d(z)\, 
\label{tosolve}
\end{equation}
we can find the initial comoving radius of our domain
(in units of the present conformal time, that is $R_i(z)/{\eta_0}$)
which it would be required to have so that its comoving size today is
equal to the comoving distance to an object at a redshift $z$.
Finally, we can calculate the radius of this domain at the present time 
$\eta_0$ and at the redshift $z$ (call it $R_{\rm max}(z)$)
with the value of $R_i(z)/\eta_0$  obtained from the previous equation.
In Fig. \ref{fig3} we plot the value of $R_{\rm max}(z) / d(z)$ as a function 
of the redshift, $z$.

If the radius of our domain at a redshift $z$ was smaller than $d(z)$ the
domain wall  would be in causal contact with us at the present time and we
could in principle detect the gravitational effect both of the domain wall and
of the different vacuum density outside our bubble.
On the other hand, if the radius of our domain at a redshift $z$ was greater
than $R_{max}$ then it would not have time to enter the sphere of radius
$d(z)$ before today. 

When the redshift of the cosmological object we
are looking at is small, that is ($z \to 0$), its comoving distance from us,
$d(z)$, is much smaller than the comoving horizon, $\eta(z)$,
at the time at which the light was emitted. Consequently, a domain 
wall with a comoving size equal to $d(z)$ at the present time would already
have a velocity very close to the speed of light by the redshift $z$.
It is easy to calculate the maximum comoving 
size, $R_{max}$, which our domain would need to have at the redshift $z$,
in order for the domain wall to enter a sphere of
comoving radius $d(z)$ centred on a nearby observer
sometime between today and redshift $z$.
This is simply given by
\begin{equation}
\frac{R_{max}(z)}{d(z)}\longrightarrow 2\, 
\label{assimp00}
\end{equation}
when $z \to 0$, because 
$d(z)$ is the distance travelled by light from a redshift $z$ until today 
(see fig. \ref{fig3}).

If we assume that the comoving radius of our bubble at a redshift $z$ is
larger than $\eta(z)$, 
then it will remain frozen in comoving coordinates until its size gets smaller
than the horizon. This means that in this case the value
of $R_{max} / d(z)$ is even smaller, approaching 
\begin{equation}
\frac{R_{max}(z)}{d(z)}\longrightarrow\frac{\alpha^{-2} \times ({\sqrt {1+4\alpha^{-n}}}-1)^{-2/n}}{2^{-2/n}}, 
\label{assimp}
\end{equation}
when $z \to \infty$ (see fig. \ref{fig3}). For a spherical domain we have
\begin{equation}
\frac{R_{max}(z)}{d_\infty} \approx 1.12\, . 
\label{assimpsph}
\end{equation}

We thus see that for the purposes of predicting the fate of the universe it
may be a plausible assumption to assume a fixed content 
in comoving coordinates. The above discussion also suggests, in
particular, that one may find {\em a posteriori} that it is
indeed a reasonable assumption if we can observe the dynamical effects of
a uniform vacuum density up to a redshift $z \ge 1$.

\section{\bf Conclusions}
\label{conc}
We have provided a simple analysis of the use of cosmological observations
to infer the state and fate of our patch of the universe. In particular,
in the same spirit of Starkman {\em et al. } \cite{STV}, we have discussed
possible criteria for inferring the present or future existence of an
inflationary epoch in our patch of the universe.

We have presented a `physical' criterion for the existence of inflation,
and contrasted it with the `mathematical' one that has been introduced
in \cite{STV}. Ours has the advantage of being able to provide (in principle)
a definite answer at the present epoch, but the disadvantage of
ultimately relying on assumptions on the content of the local universe and on field dynamics. We consider our
assumptions to be plausible, but we can certainly conceive of (arguably
contrived or fine-tuned) mechanisms that would be capable of violating it.

\acknowledgments
C. M. is funded by FCT (Portugal) under  `Programa PRAXIS XXI' (grant
no. PRAXIS XXI/BPD/11769/97). We thank Centro de Astrof\'{\i}sica da
Universidade do Porto (CAUP) for the facilities provided.

\end{document}